\documentclass{article}
\usepackage{spconf,amsmath,graphicx}
\usepackage{booktabs}
\usepackage{color}
\usepackage{comment}
\usepackage{scalerel}
\usepackage{upgreek}
\usepackage{bbm}
\usepackage{amssymb}
\usepackage{csquotes}

\let\lctau\tau 
\renewcommand{\tau}{\scalerel*{\lctau}{X}}
\title{Self-supervised learning of audio representations using angular contrastive loss}
\name{Shanshan Wang, Soumya Tripathy, Annamaria Mesaros\thanks{This work was supported in part by Academy of Finland grant 332063 ``Teaching machines to listen". The authors wish to thank CSC-IT Centre of Science Ltd., Finland,  for providing computational resources.}}
\address{Computing Sciences, Tampere University,
Tampere, Finland
}

\begin{document}
%
\maketitle
\begin{abstract}

In Self-Supervised Learning (SSL), various pretext tasks are designed for learning feature representations through contrastive loss. However, previous studies have shown that this loss is less tolerant to semantically similar samples due to the inherent defect of instance discrimination objectives, which may harm the quality of learned feature embeddings used in downstream tasks. To improve the discriminative ability of feature embeddings in SSL, we propose a new loss function called Angular Contrastive Loss (ACL), a linear combination of angular margin and contrastive loss. ACL improves contrastive learning by explicitly adding an angular margin between positive and negative augmented pairs in SSL. Experimental results show that using ACL for both supervised and unsupervised learning significantly improves performance. We validated our new loss function using the FSDnoisy18k dataset, where we achieved 73.6\% and 77.1\% accuracy in sound event classification using supervised and self-supervised learning, respectively.
\end{abstract}
\section{Introduction}
\label{sec:intro}
Supervised learning methods have achieved significant performance in solving various tasks like image classification \cite{karpathy2014large, he2016deep, dosovitskiy2021an} and sound event detection \cite{heittola2013context, cakir2017convolutional, 9524590}. These achievements are not possible without the considerable effort spent collecting the dataset and manually labeling the classes, making the training process costly and time-consuming. Although deep architectures with supervision achieve superior accuracy, the commonly used Cross-Entropy (CE) loss does not explicitly optimize the feature embedding space. CE loss fails to learn compact intra-class clusters with clear boundaries between them \cite{choi2020amc}, affecting the interpretability and quality of features learned in the supervised setups. Many auxiliary loss functions are added to the CE loss to address the feature interpretability issues. For example in \cite{choi2020amc}, angular margin contrastive loss, a geometric constraint, was added to help separate the class features by a predefined margin. The quality of the features slightly improves from these constraints in supervised setups, giving a little improvement over CE loss.

In the recent years, large amounts of audio-visual materials are uploaded in the social media every day. This leads to design of many unsupervised and self-supervised learning algorithms to learn features from large amount of data without any human supervision.
Chen et al. \cite{chen2020simple} adopted the contrastive learning as a pretext task to learn  robust image features in a self-supervised setup. These features can then be used in  downstream tasks like image classification. The contrastive loss learns the features by maximizing the feature agreement between the \textit{positive pairs} (augmentation of same image) and minimizing the agreement for the \textit{negative pairs} (augmentation of different images). Due to its success in the vision community, contrastive learning has now gained a lot of attention in audio feature learning \cite{saeed2021contrastive, fonseca2021unsupervised, morgado2020learning}. Fonseca et al. \cite{fonseca2021unsupervised} applied the same idea to learn sound event representations. Another often applied approach to contrastive learning is correspondence between modalities. For example Morgado et al. \cite{morgado2020learning} used audio-visual correspondence and spatial alignment between 360\textdegree \ video and Ambisonics audio, while Wang et al. \cite{wang2022self} further investigated on the object-oriented video crops and Ambisonics alignment.

Despite its success, contrastive learning has some underlying problems, as pointed out in a thorough analysis by Wang et al. \cite{wang2021understanding}. They find that while learning features using contrastive loss, there exists a trade-off between \emph{uniformity} of the features: feature embeddings should be uniformly distributed on unit hypersphere \cite{wang2020hypersphere} and \emph{tolerance} of the features: features of semantically similar items should be close to each other. The cross-entropy loss used in contrastive learning only focuses on bringing the features of the same instances together while pushing the features of different instances away, without focusing on the semantical relationships between them. Although moving semantically similar items away from each other can generate uniform feature distribution, they may lead to low tolerance and poor features for downstream tasks like classification. In \cite{wang2021understanding}, they find that the temperature parameter in the contrastive loss controls this trade-off, and either extreme of this parameter value leads to poor downstream task performance. This encourages us to design a better loss function that can increase the tolerance i.e separation between semantically dissimilar items and increase the downstream task performance, unlike the the traditional contrastive loss.

In this paper, we propose a new loss, ACL, for SSL that adds angular margin loss to the instance discrimination task in contrastive learning. We perform experiments on supervised and self-supervised learning using the ACL and achieve higher performance than the cross-entropy-based loss functions. Finally, we analyze why the ACL improves the downstream classification accuracy with the help of feature tolerance and uniformity introduced in \cite{wang2021understanding}.

\section{Self-supervised learning using ACL}
\label{sec:method}

\begin{figure}
    \centering
    \includegraphics[width=\linewidth]{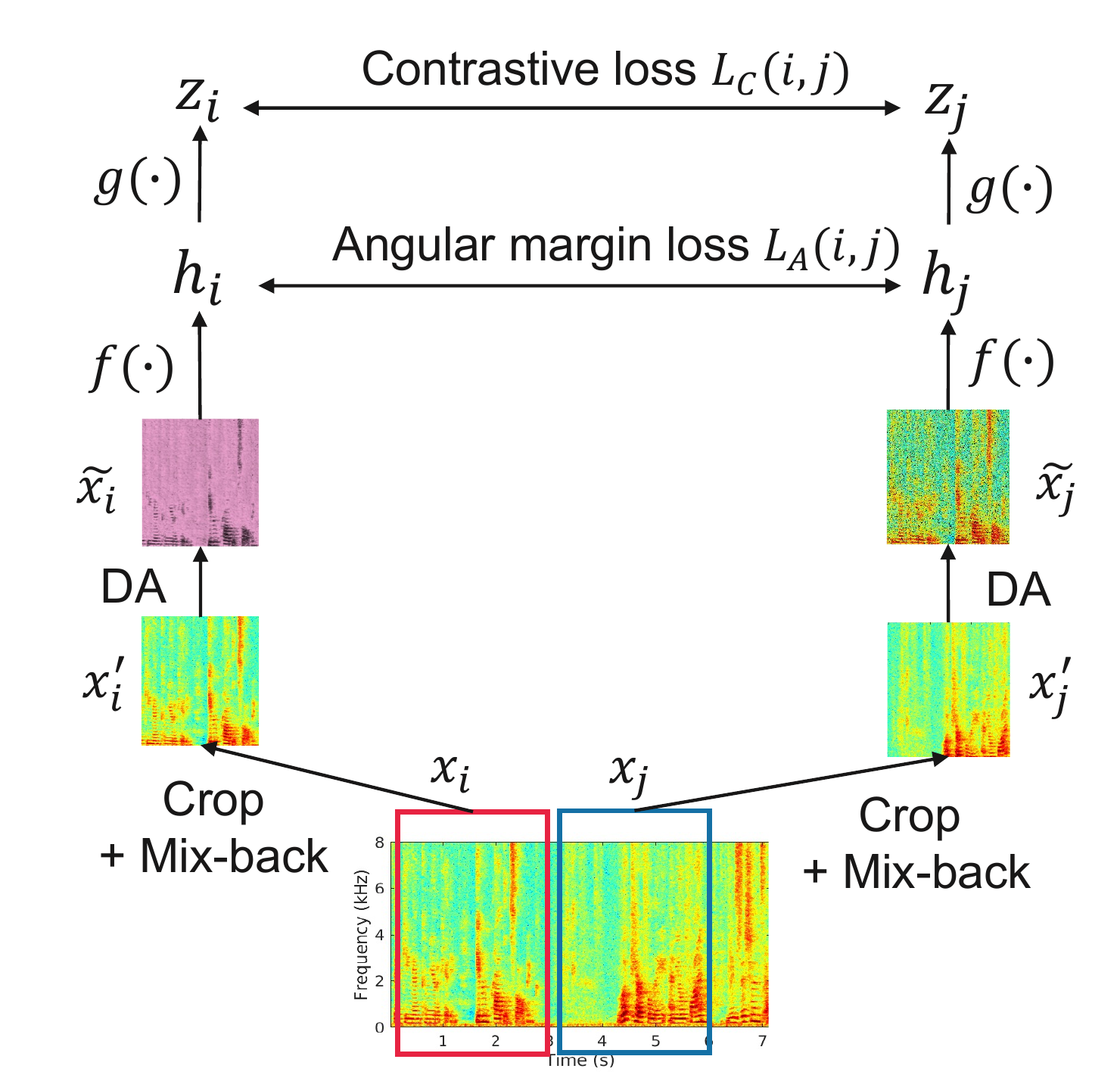}
    \caption{The main diagram of the proposed method. The total loss is the linear combination of the contrastive loss $L_C$ and the angular margin loss $L_A$. DA stands for data augmentation, $f(\cdot)$ for encoder network, and $g(\cdot)$ for the projection head. With the exception of the angular margin loss applied between $h_i$ and $h_j$, the setup strictly follows \cite{fonseca2021unsupervised} for the purpose of comparison.}
    \label{fig:main_concept}
    \vspace{-10pt}
\end{figure}

Our self-supervised audio feature learning framework is illustrated in Fig. \ref{fig:main_concept}. For the pretext task, we first represent each audio clip by its log-mel spectrogram and then randomly crop two time-frequency (TF) patches $x_i$ and $x_j$ from each clip. Mix-back operation \cite{fonseca2021unsupervised} which mixes the current patch with a background patch drawn from training set is applied to get $x_i'$ and $x_j'$. Then we apply data augmentation techniques such as random sized cropping \cite{takahashi2019data}, specAugment \cite{park19e_interspeech}, and Gaussian blurring to get $\tilde{x_i}$ and $\tilde{x_j}$. Our final task is to distinguish that in a batch $x_i$ corresponds $x_j$. We employ ResNet18 \cite{he2016deep} as the encoder network to extract the feature embeddings $h_i=f(\tilde{x_i})$ and $h_j=f(\tilde{x_j})$. Finally as suggested in \cite{chen2020simple}, feature embeddings are projected to a lower dimension features $z_i=g(h_i)$ and $z_j=g(h_j)$ by a projection head consisting of an MLP with one hidden layer, batch normalization and a non-linear ReLU.

For the correspondence learning in a contrastive way, we utilize the NT-Xent loss \cite{chen2020simple}, defined as
\begin{equation}
    L_C(i,j) = -\textit{log}\frac{\textit{exp}(\textit{sim}(z_i,z_j)/\footnotesize{\tau})}{\sum_{k=1}^{2N} \mathbbm{1}_{[k \neq i]} \textit{exp}(\textit{sim}(z_i,z_k)/\footnotesize{\tau})},
\end{equation}
where \textit{sim($\cdot$)} calculates the cosine similarity between features $z_i$ and $z_j$, $\mathbbm{1}_{[k \neq i]} \in \{0,1\}$ is an indicator function being $1$ for $k\neq i$ and $0$ for $k=i$, and $\footnotesize{\tau}$ is the temperature scaling factor which controls how the negative samples are penalized, with smaller value penalizing more and vice versa. Essentially, this is a temperature scaled cross-entropy loss that pulls the feature representations of positive pairs closer while pushing away those from negative pairs. As suggested in \cite{wang2021understanding}, in the above loss function there exist a trade-off between uniformity (embeddings should be uniformly distributed on unit hypersphere) and tolerance (features of semantically similar items should be close to each other). The cross-entropy based contrastive loss inherently achieves more uniformity and fails to add clear inter-class margins \cite{wang2020hypersphere, wang2021understanding}, which may be more desirable for downstream tasks like classifications. Moreover, the trade-off in the contrastive loss is controlled by the temperature parameter $\footnotesize{\tau}$. In \cite{wang2021understanding} they experimentally show that increasing $\footnotesize{\tau}$ from zero to one, correlates negatively with uniformity and positively with tolerance. Although we expect that higher temperature can lead to better tolerance, therefore better downstream task performance, the loss of uniformity hinders the performance significantly. In order to avoid this trade-off, we apply angular margin loss directly on the features $h_i$ and $h_j$ as defined:
\begin{equation}
L_A(i,j) = \begin{cases}
(\cos ^{ - 1} \langle h_i,h_j \rangle)^{2} &\text{if $S_{ij}=1$}\\
\textit{max}(0,m_g-\cos ^{ - 1} \langle h_i,h_j \rangle)^2 &\text{if $S_{ij}=0$}
\end{cases},
\end{equation}

\noindent where $m_g > 0$ is the angular margin, $S_{ij}$ evaluates to $1$ for the positive pairs where the angular distance is minimized; $S_{ij}$ evaluates to $0$ for the negative pairs, ensuring a distance margin $m_g$ \cite{choi2020amc}. It is to be noted that unlike contrastive loss, we apply the angular margin loss on $h_i$ and $h_j$. We believe that geometric constraints are retained better without the lower dimensional projection heads. Our experimental validation also points to the same conclusion.

Finally, we combine both losses to form $ACL$ defined below as:
\begin{equation} \label{eq:combined_loss}
    ACL = \alpha * L_C + (1-\alpha) * L_A,
\end{equation}
The $\alpha \in [0,1]$ is the weighting parameter. Setting $\alpha = 1$, makes ACL same as contrastive loss $L_C$, while making $\alpha=0$ reduces it to $L_A$. Different values of $\alpha$ are investigated throughout the experiments.

\begin{figure*}
    \centering
    \includegraphics[width=\linewidth]{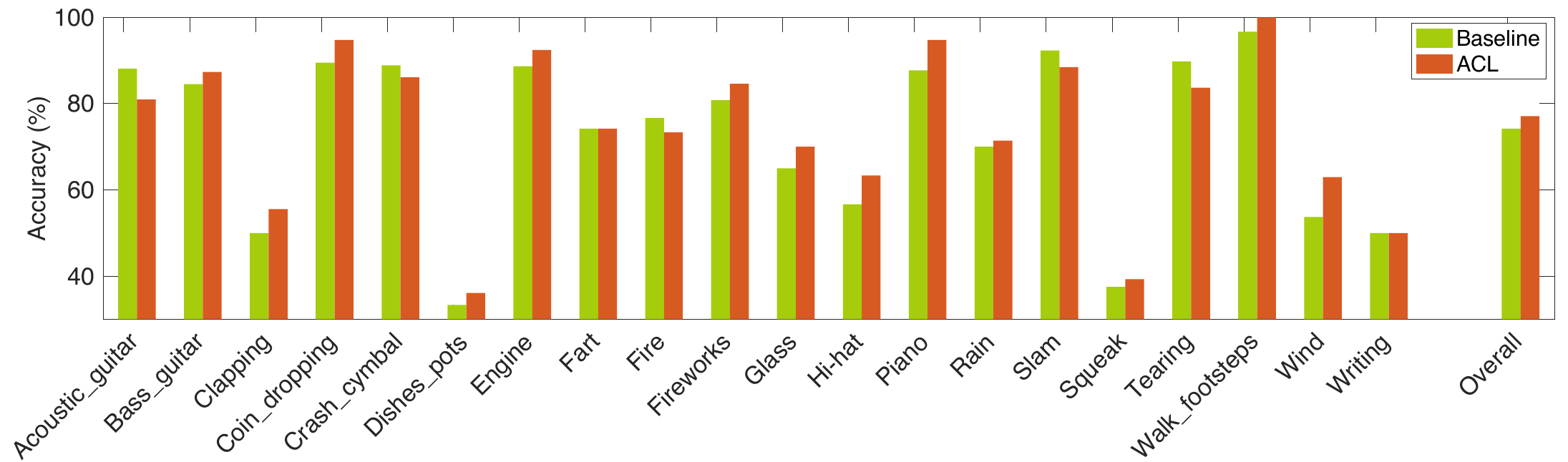}
    \vspace{-10pt}
    \caption{Class-wise accuracies (\%) for in-domain features evaluation. Green bar shows the baseline results trained on contrastive loss only and red bar displays our proposed ACL with $\alpha = 0.3$.}
    \label{fig:bar_plot}
\end{figure*}

\section{Dataset}
\label{sec:data}
We perform the experiments on FSDnoisy18k dataset \cite{fonseca2019learning}, containing 42.5 hours of audio belonging to 20 classes drawn from AudioSet Ontology \cite{gemmeke2017audio}. The length of the audio clips is  between 0.3 and 30 seconds. The dataset is provided with a predefined training and testing set containing 17585 and 947 clips respectively. The training set is further divided into a small clean set of 1772 clips containing correct and complete labels, and a larger noisy set of 15813 clips that received no human validation, and are categorized using only the user-provided tags.

\section{Experimental setup and results}
\label{sec:setup}

We perform experiments in supervised and self-supervised learning scenarios to investigate the effect of ACL. In supervised learning, we use the sound event labels from the dataset to perform sound event classification. As a baseline, we first train the system of Fonseca et al. \cite{fonseca2019learning} by using a standard cross-entropy objective.  To evaluate ACL in this setup, we then add the angular margin loss at the second-last layer, right before the linear classification layer, as in \cite{choi2020amc}. The final loss function is the linear combination of CE and angular margin loss with a factor $\alpha$, similar to Eq. \ref{eq:combined_loss}.

For SSL, we learn the audio features from the dataset by solving the pretext task explained in Sec. \ref{sec:method}. The feature learning step does not require any labels available in the dataset. As the baseline, we first reproduce the system in \cite{fonseca2021unsupervised} using traditional contrastive loss and then train the same system with our ACL for pair comparison of performances. For both experiments we keep the temperature parameter $\footnotesize{\tau} =  0.2$ as given in \cite{fonseca2021unsupervised}. After the feature learning step, we evaluate the quality of the features by testing them on the downstream task of sound event classification. Codes are available at here \footnote{https://github.com/shanwangshan/Self\_supervised\_ACL}.

\begin{figure*}
    \centering
    \includegraphics[width=\linewidth]{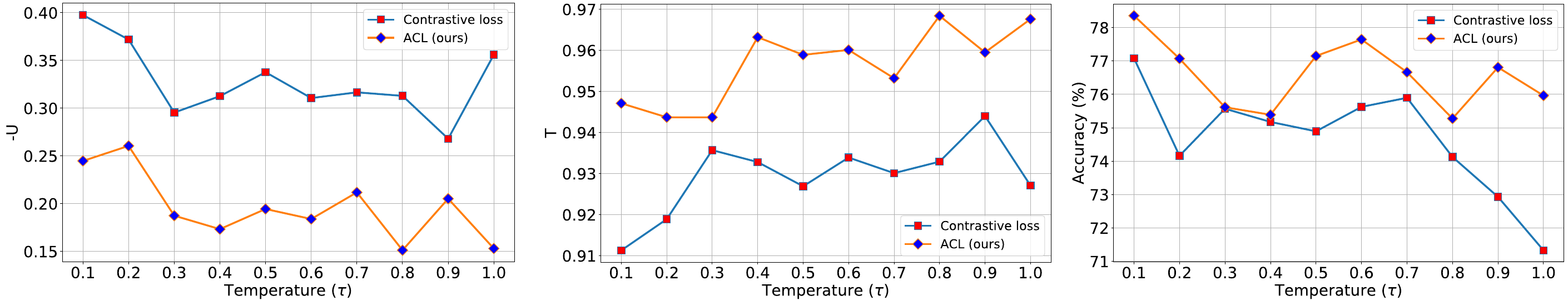}
    \vspace{-20pt}
    \caption{Uniformity, tolerance and downstream classification accuracies with varying temperature for contrastive loss and ACL.}
    \label{fig:uniformity_tolerance}
\end{figure*}

\subsection{Supervised learning}
Supervised learning results are listed in Table \ref{tab:supervised_results} for different subsets of FSDnoisy18k used for training the models. We indicate the $\alpha$ value in the ACL as in Eq. \ref{eq:combined_loss} corresponding to the best validation performance in the last column of Table \ref{tab:supervised_results}. It is evident that irrespective of the data subset used in training, the performance significantly improves by adding margin loss: $3.5\%$ improvement for the whole training data, $1.9\%$ for the Noisy subset, $6.2\%$ for the Noisy small subset, and $4.3\%$ for the Clean subset. The improvements in the results are tested and found statistically significant based on 95\% confidence intervals.

Based on all our experiments, we noticed that for clean data, $\alpha=0.5$ gives the best accuracy. When the labels are reliable, CE loss and angular margin loss should be given equal weights to achieve superior performance. As for noisy data, we notice that a smaller $\alpha$, between $[0.2,0.4]$ gives the best performance. Giving a smaller weight to CE loss implies that we do not trust the labels enough to penalize the system based on the information provided only by the binary targets. This seems intuitive, since data points with wrong labels may result in very high loss values, influencing the training process strongly. Giving more weights to angular margin loss seems to mitigate the effect of label noise in the dataset.

\label{sec:results}


\begin{table}[]
    \centering
    \begin{tabular}{l|c|c|c}
    \toprule
    FSDnoisy18k  & Supervised  & CE &  with $\alpha$ \\
    subsets& baseline \cite{fonseca2019learning} &+ margin loss & \\
    \midrule
    All  &  70.1 & \textbf{73.6} & 0.7 \\
    Noisy & 66.4  & \textbf{68.3} & 0.4  \\
    Noisy small & 33.7  & \textbf{39.9} & 0.2  \\
    Clean & 60.2  & \textbf{64.5}  & 0.5  \\
    \bottomrule
    \end{tabular}
    \caption{Test classification accuracies (\%) of supervised learning approaches trained on different subsets of FSDnoisy18K.}
   \label{tab:supervised_results}
   \vspace{-4pt}
\end{table}

\subsection{Self-supervised learning}

We test the self-supervised features from baseline contrastive loss and our ACL on in-domain downstream tasks of sound event classification. For this purpose, we add an extra linear classifier on top of the feature extraction layer for both  methods and obtain $74.2\%$ average classification accuracy for baseline and 77.1\% for our ACL model, with an improvement of 2.9\% (statistically significant). We provide class-wise classification accuracy on the test set for both the systems in Fig. \ref{fig:bar_plot}. It can be observed that 13 out of 20 classes benefit from ACL, particularly the ones with lowest performance in the baseline.

Table \ref{tab:my_label} summarizes the performances obtained for the FSDnoisy18k test set concerning different training methods. Self-supervised learning is superior to the supervised learning method, even given the small data size, which is also pointed out in Fonseca et al. \cite{fonseca2021unsupervised}. For ACL, we notice that its improvement in supervised learning also holds for self-supervised learning. ACL brings a $3.5\%$ improvement for supervised learning and a $2.9\%$ increase for SSL comparison to its baselines. Moreover, similar to the supervised case, a moderate $\alpha$ value ($0.3$ for SSL) gives the optimal performance. While the best $\alpha$ value can be selected based on the validation set, the presented results suggest that choosing a value between  $[0.2, 0.4]$ is generally suitable.

\begin{table}[]
    \centering
    \begin{tabular}{l|c}
    \toprule
         Training method & Best performance  \\
         \midrule
         Supervised baseline & 70.1 \\
         Supervised with ACL & \textbf{73.6} \\
         SSL baseline & 74.2 \\
         SSL with ACL & \textbf{77.1} \\
         \bottomrule
    \end{tabular}
    \caption{Downstream classification accuracy (\%) obtained on the FSDnoisy18k test set with the different training methods, the systems are trained on the complete training set. }
    \label{tab:my_label}
  \end{table}
  \vspace{-5pt}

\subsection{ACL features analysis}
We compare the properties of the feature embeddings learned through contrastive loss and ACL to find a suitable explanation for the improved performance using ACL. To understand the feature quality, we calculate uniformity (\(U\)) and tolerance (\(T\)) as defined in \cite{wang2021understanding, wang2020hypersphere} and compare its trend to the downstream task's accuracies. By randomly sampling feature vectors from the test set, we calculate these two metrics as  in \cite{wang2021understanding}:
{\small
\begin{align*}
  U(z; t)  = \textit{log} \ \displaystyle \mathop{\mathbb{E}}_{z_x, z_y \sim data} \left[\textit{exp}\left(-t \|z_x- z_y\|_2^2 \right) \right], \ t > 0\\
  T = \displaystyle \mathop{\mathbb{E}}_{z_x, z_y \sim data} \left[z_x^Tz_y \quad \mathbbm{1}_{[l(z_x) = l(z_y)]} \right],
\end{align*}}%
where \(l(\cdot)\) represents the class labels attached to the feature vector \(z\) and $\mathbbm{1} \in \{0,1\}$ is an indicator function being \(1\) if both features have same class labels. We plot these two metrics along with average downstream task accuracy by varying temperature values. From Fig. \ref{fig:uniformity_tolerance}, it can be seen that irrespective of the temperature value, our ACL scores slightly higher feature tolerance with the degradation of feature uniformity. We also notice that ACL scores higher accuracy than the baseline contrastive loss in downstream classification task for all the temperatures. This slight increase in the feature tolerance value seems to be more beneficial and the degradation in the uniformity does not necessarily harm the feature quality for our downstream tasks.
As we understand from \cite{wang2021understanding}, the tolerance points to the existence of local clusters of semantically similar objects. So our ACL creates better inter-class separation than NT-Xent loss, thereby achieving higher accuracy in downstream tasks. Moreover, from this experiment, it is safe to say that ACL leads to stable class accuracy concerning temperature parameters by relaxing the trade-off between uniformity and tolerance introduced from NT-Xent contrastive training.

\section{Conclusions}
\label{sec:concl}
We study the drawbacks of the NT-Xent contrastive loss and propose angular contrastive loss to improve the feature quality in the SSL training. We validate ACL audio representation learning through both supervised and SSL training. Experimental results suggest that classification type of downstream tasks benefit from adding ACL in SSL. Future work will be directed towards training with larger and possibly multimodal datasets, e.g. audio-visual correspondence or alignment pretext tasks for a more thorough analysis of its benefits to a larger variety of data and downstream tasks.

\bibliographystyle{IEEEbib}
\bibliography{references}

\end{document}